# A data mining-based solution for detecting suspicious money laundering cases in an investment bank

Nhien An Le Khac
School of Computer Science
University College Dublin
Belfield, Dublin 4, Ireland
e-mail: an.lekhac@ucd.ie

Sammer Markos
School of Computer Science
University College Dublin
Belfield, Dublin 4, Ireland
e-mail: sammer.markos@ucd.ie

M-Tahar Kechadi
School of Computer Science
University College Dublin
Belfield, Dublin 4, Ireland
e-mail: tahar.kechadi@ucd.ie

*Abstract*—Today, money laundering poses a serious threat not only to financial institutions but also to the nation. This criminal activity is becoming more and more sophisticated and seems to have moved from the cliché of drug trafficking to financing terrorism and surely not forgetting personal gain. Most international financial institutions have been implementing anti-money laundering solutions to fight investment fraud. However, traditional investigative techniques consume numerous man-hours. Recently, data mining approaches have been developed and are considered as well-suited techniques for detecting money laundering activities. Within the scope of a collaboration project for the purpose of developing a new solution for the anti-money laundering Units in an international investment bank, we proposed a simple and efficient data mining-based solution for anti-money laundering. In this paper, we present this solution developed as a tool and show some preliminary experiment results with real transaction datasets.

## I. Introduction

Money laundering (ML) is a process to make illegitimate income appear legitimate; this is also the process by which criminals attempt to conceal the true origin and ownership of the proceeds of their criminal activity. ML has been defined by Genzman as an activity that "knowingly engage in a financial transaction with the proceeds of some unlawful activity with the intent of promoting or carrying on that unlawful activity or to conceal or disguise the nature location, source, ownership, or control of these proceeds" [1]. Through money laundering, criminals try to convert monetary proceeds derived from illicit activities into "clean" funds using a legal medium such as large investment or pension funds hosted in retail or investment banks. This type of criminal activity is getting more and more sophisticated and seems to have moved from the cliché of drug trafficking to financing terrorism and surely not forgetting personal gain. Today, ML is the third largest "Business" in the world following the Currency Exchange and the Automobile Industry. According to the United Nations Office on Drug and Crime, the worldwide value of laundered money in one year ranges from $500 billion to $1 trillion [2] and from this approximately $400-450 billion is associated with drug trafficking. The rest is from other forms of organised crime such as fraud, robbery, forgery & counterfeiting, blackmail, extortion and terrorist activity. These figures are at times modest and are partially fabricated using statistical models, as nobody exactly knows the true value of money laundering, one can only forecast according to the fraud that has already been exposed. Nowadays, it poses a serious threat not only to financial institutions but also to the nation. Some risks faced by financial institutions can be listed as reputation risk, operational risk, concentration risk and legal risk. At the society level, ML could provide the fuel for drug dealers, terrorists, arms dealers and other criminals to operate and expand their criminal enterprises. Briefly, nations care about ML because they care about their political and economic stability. Hence, the governments, financial regulators require financial institutions to implement processes and procedures to prevent/detect money laundering as well as the financing of terrorism and other illicit activities that money launderers are involved in. Therefore, anti-money laundering (AML) is of critical significance to national financial stability and international security.

Traditional approaches to AML followed a labour-intensive manual approach because ML is a sophisticated activity with many way of laundering money. These approaches can be classified into the identification of money laundering incidences, detection, avoidance and surveillance of money laundering activities [3]. Indeed, given that the volume of banking data and transactions have increased in number of ways, such approaches need to be supported by automated tools for detecting money laundering's pattern. Meanwhile, AML software tools in the market are normally rule-based that make the decisions using some sets of predefined rules and thresholds based on mean and standard deviation values.

Besides, data mining techniques (DM) [4] have been proven to be well suited for identifying trends and patterns in large datasets. Therefore, DM techniques are expected to be applied successfully in the area of AML. Nevertheless, there is still little research concerning this bias especially a DM framework/solution for supporting AML experts in their daily tasks. Recently, there are some AML approaches based on DM that have been proposed and discussed in literature. Most of these approaches try to recognize ML patterns by different techniques such as support vector machine [5], correlation analysis [6], histogram analysis [6], etc. They aim to provide techniques for detecting a variety of ML by exploring a massive dimensionality of datasets including customers x accounts x products x geography x time. However, these approaches are more or less appropriate for the cash world and not scaled well for investment activities due to the lack of good methods in choosing parameters and they still have performance issues. In our recent work [7][8][16], we proposed a method to identify customers and analysed important parameters linked to ML patterns in an international investment bank. Customer behaviour in investment activities is complicated because it is influenced by many factors. We also show that by choosing suitable parameters, simple DM techniques can be applied together to detect suspicious ML cases in investment activities. In this paper, we present a DM framework that bases on a combination of clustering and classification techniques for analysing transaction datasets to detect these cases.

The rest of this paper is organised as follows: Section II deals with recent works on this subject. Section III resumes our DM framework for detecting money-laundering activities. We present our approach for analysing transaction datasets in Section IV. Experiment results of this approach are presented and discussed in Section V. Finally, we conclude in Section VI.

## II. RELATED WORKS

[6] applied a discretisation process on their datasets to build clusters. They map their feature space "customer x time x transaction" to $n+2$ dimensional Euclide space: $n$ customer dimensions, $1$ time dimension and $1$ transaction dimension. They firstly discretise the whole timeline into difference time instances. Hence, each transaction is viewed as a node in one-dimensional timeline space. They project all transactions of customers to the timeline axis by accumulating transactions and transaction frequency to form a histogram. They create clusters based on segments in the histogram. A local and a global correlation analysing are then applied to detect suspicious patterns. This approach improves firstly the complexity by reducing the clustering problem to a segmentation problem [9]. Furthermore, it is more or less appropriate for analysing individual behaviours or group behaviours by their transactions to detect suspicious behaviours related to "abnormal" hills in their histogram. However, as we have to analyse many customers with many transactions with a variety of amounts for a long period, it is difficult to detect suspicious cases, as there are very few or no "peak hills" in the histogram. Firstly, another global analysis is needed and we can then apply this method for further analysis in this case.

Another approach for AML is using support vector machine (SVM) [10]. In [11], authors propose an extension of SVM to detect unusual customer behaviour. They present a combination of an improved RBF kernel [12] with the definition of distinct distant [13] and supervised/unsupervised SVM algorithms (C-SVM, one-class SVM). One-class SVM [10] is an unsupervised learning approach used to detect outliers based on unlabeled training datasets which is highly suitable for ML training sets. The advantage of this approach is that it can deal with heterogeneous datasets. However, there is a performance issue due to the lack of parameter selection.

## III. DATA MINING FRAMEWORK FOR AML

A framework for detecting ML activities is normally consisted of four layers [14][15] corresponding to four levels of analysis: transaction, account, institution and multi-institution. The first three levels: transaction, account and institution are the most important where the last one depends more or less on the organisations and their policy. Our DM framework also includes these three levels and is consist of different components as shown in the Fig.1.

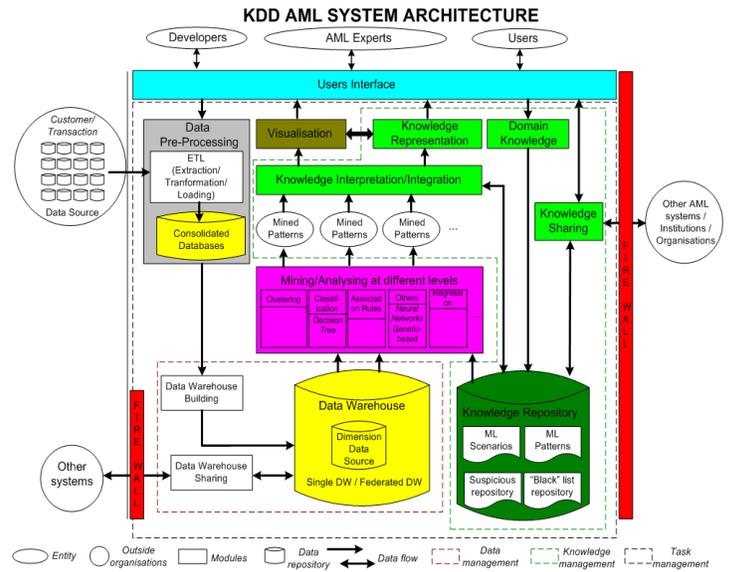

Figure 1. DM Architecture for detecting AML.

### A. Data Pre-Processing

The main role of this component is to extract and clean raw datasets from data sources located in different sites of this international bank. It then integrates them into consolidated databases that are used to build a data warehouse of customer information and customer transactions. One of the most challenges in this step is the data quality issue. In banking and finance, datasets has a different set of quality problems at the instance level and in our case, most of these problems relate to customer information. Some of them can be listed as:

- *missing values*, dummy values or null. This occurs in most of the data fields in all databases analysed except the customer identification (ID), the customer type (corporate, individual and joint) and the fund name;
- *misspellings*, usually typos and phonetic errors.

Furthermore, banking datasets are normally managed in distributed way for the flexibility and security reasons. The independence and heterogeneity of each data source can also be data quality issues especially when an integrating task is required, as all conflicts must be solved. Some basic data quality issues are solved by this pre-processing component. Other issues related to customer information are implemented in customer identification module [7] integrated in our knowledge management component.

Data quality issue related to customer transactions is normally at the logic level, e.g. the mapping error where transactions of a customer are copy from one system to another system and sometimes not all of transactions are copied. Consequently, we do not have a whole picture of his transactions but its snapshot.

### B. Data Mining

This component provides classification and clustering techniques for the most basic level of this framework: analysing transaction datasets. At this level, transaction records are extracted for investigations. However, they have a few analytical contexts because they do not provide links to accounts or other data. In the second level, multiple transactions are associated with specific accounts. Aggregation of transactions with individual accounts gives a general view of these accounts about their financial activity. In our approach, customer behaviours have been detected by mining techniques such as: clustering that groups similar transactions and builds suspicious profiles; classification that classify customers into pre-defined categories of risk. Details on these techniques will be found Section IV.

### C. Knowledge Management

Results of mining process, experience of AML experts, running results are collected, stored in relevant repositories and analysed by this component. It also generates significant, interpretable rules and knowledge. This component controls all the data mining process by proposing different strategies mining as well as for integrating and coordinating to achieve the better performance. For example, it integrates clustering results and learning results to build a decision tree for detecting suspicious case. Its architecture also allows to share knowledge to other systems with regards to their privacy policy.

## IV. TRANSACTION MONITORING

In this section, we present our solution for analysing investment transactions. This solution is implemented in the data mining and knowledge components of our framework. As mentioned above, transactions and accounts cannot be separately investigated; they should be aggregated to give a general view of customers' behaviour. Normally, this analysis is based on two important characteristics: frequency of transactions and the value of each transaction.

Current solutions apply these two characteristics in a set of rules to detect suspicious cases. Most of the vendor software approaches found in the market are based on a decision tree using frequency and value of transactions as a marker, the thresholds for these markers are based on averages and the standard deviation. This approach only uses one-way comparison i.e. customer X's behaviour against customer X's previous "normal" behaviour. This approach is reasonably adequate for the cash world (accounts). However, they are not efficient for the investment market because there are many factors that influence the frequency of trades in investment banking such as political environment, market climate, fund prices, currency exchange rates, etc. [8] also give an example of the variety of transaction frequencies that exists among different investment funds.

Briefly, an efficient solution to investigate ML in investment banking is to determine relevant parameters to decrease the number of dimensions (attributes) and to improve performance. In our recent work [8], by analysing customer transactions, we proposed parameters that were used in detecting suspicious cases. Concretely, we defined two parameters: $\Delta_1$, the proportion between the redemption value and the subscription value conditional on time (daily, weekly, monthly etc) and $\Delta_2$, the proportion between a specific redemption value and the total value of the investors' shares conditional on time as below:

$$\Delta_1 = \begin{cases} \left|\frac{\alpha_i}{\beta_i}\right|_{\tau_i} & \text{Where } \alpha_i \leq \beta_i \\ \left|\frac{\beta_i}{\alpha_i}\right|_{\tau_i} & \text{Otherwise} \end{cases} \qquad \Delta_2 = \left|\frac{\beta_i}{\theta_i}\right|_{\tau_i} \quad (1)$$

where $\alpha_i$ is the subscription value and $\alpha_i \in [0…+\infty]$, $\beta_i$ is the redemption value and $\beta_i \in (0…+\infty]$, $\theta_i$ is the value of the investors shares and $\theta_i \in [0…+\infty]$, $\tau_i$ is time and $\tau_i \in$ [Day, Week, 1 month, 3 months, 6 months or 12 months] depending on the investigating level (by day, by week, etc.). Note that the value of the transactions (subscription or redemption) of each investor in an investment fund is aggregated by time: daily, weekly, monthly, 3 monthly, 6-monthly and yearly, for example $\alpha_i$ is the subscription value of the day $\tau_i$ if the investigating level is by day. Another example is shown in the Table I, in week 30, investor A01 had a total subscription of $100 and had a total redemption of $90. His $\Delta_1$ and $\Delta_2$ are 0.9 and 0.82 respectively.

Through the evaluation of these parameters on real transaction datasets, we recognise that the parameter $\Delta_1$ reflects only the relationship between subscription and redemption amount on the time when these activities happen, i.e. subscriptions and redemptions are carried out in the same day or in the same week, etc. Besides, from the AML experts' experience, the relevant subscriptions of a redemption activity in suspicious cases are normally not only in the current investigation term (week, month...) but also in its short previous term (two, three weeks or two three month ago).

For example, a redemption is carried out today and its related subscriptions could be carried out in the last few days, and then if we consider these two activities together, it would be a suspicious case.

TABLE I. AN EXAMPLE OF Δ1 AND Δ2 OF INVESTORS IN FUND A BY WEEK

| ID | Time | Sub[a] | Red[b] | Values of the investor shares | Δ1 | Δ2 |
|---|---|---|---|---|---|---|
| ... | ... | ... | ... | ... | ... | ... |
| A01 | Week 30 | 100 | 90 | 110 | 0.9 | 0.82 |
| A02 | Week 30 | 50 | 20 | 300 | 0.4 | 0.06 |
| A01 | Week 33 | 100 | 90 | 120 | 0.9 | 0.75 |
| A02 | Week 37 | 100 | 80 | 380 | 0.8 | 0.21 |
| A03 | Week 37 | 500 | 400 | 900 | 0.8 | 0.44 |
| A04 | Week 50 | 700 | 300 | 1500 | 0.43 | 0.2 |
| ... | ... | ... | ... | ... | ... | ... |

a. Subscription, b. Redemption

In order to tackle this problem, we refine the definition of $\Delta_1$ as below:

$$\Delta_1 = \begin{cases} \left|\dfrac{\alpha_l}{\beta_j}\right|_{\tau_j} & \text{Where } \alpha_l \leq \beta_j \\ \left|\dfrac{\beta_j}{\alpha_l}\right|_{\tau_j} & \text{Otherwise} \end{cases} \quad (2)$$

where $\alpha_l = \max(\alpha_i): i \in [j-k, j]$, $\forall j >= k$, $\alpha_i$ is the subscription value and $\alpha_i \in [0...+\infty]$, $\beta_j$ is the redemption value and $\beta_j \in (0...+\infty]$, $\tau_j$ is time as in the definition (1). Moreover, The parameter $k$ can be chosen by AML experts. It normally varies from 3 to 5. For instance, in the Table I, $\Delta_1$ of the customer A01 at Week 33 is not the proportion between the redemption value and the subscription value in Week 33 but now is the maximum of subscription values from Week 30 to Week 33 ($k = 3$ in this case).

Fig.2 shows the whole AML process of our approach. In the next step (Fig.2: step (4)), we apply a clustering technique for each $\Delta_1$ and $\Delta_2$ at two levels: fund and investor. A centre-based technique is chosen because the shape of cluster (convex) does not really affect on the final decision. There are two types clustering: one is carried out on the whole datasets and another on a suspicious screening of datasets depending on the datasets size. The reason is that if the datasets size is large, the first clustering type includes a repetitiveness of a clustering algorithm on transaction datasets to determine the suspicious group. This is a time-consuming step. We need, moreover, interaction with AML experts at each loop of this stage. Consequently, this step affects the overall performance of our solution in the term of running time.

Fig.3, for instance, shows four clusters of fund SK[1] datasets (~70000 elements) based on two variables $\Delta_1$ and $\Delta_2$

---
[1] Real name of fund can not be disclosed because of confidential agreement of the project

after the first running of a centre-based clustering algorithm. Generally, the most suspicious cases should obtain high values in $\Delta_1$ and $\Delta_2$. In this example, Cluster 3 contains not only elements with high values in $\Delta_1$ and $\Delta_2$ but also the elements with low values in $\Delta_2$. Hence, this clustering algorithm is required to perform several times on this cluster 1 and its subsets to determine the suspicious group.

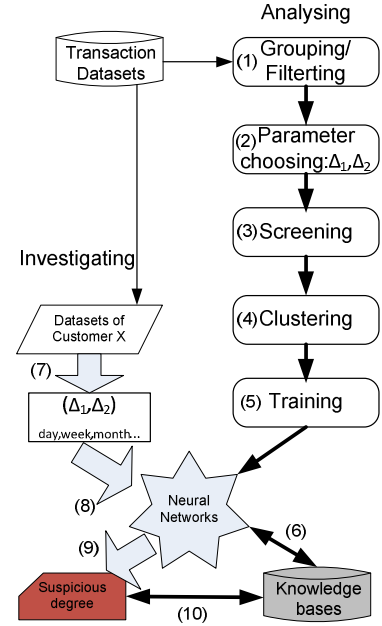

Figure 2. AML analysing and investigating process.

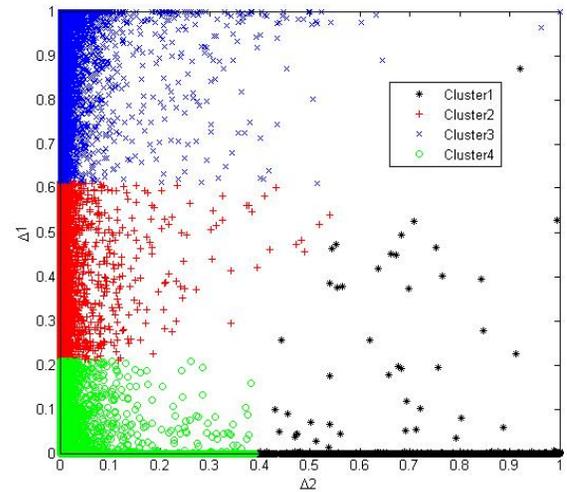

Figure 3. Clustering of fund SK datasets based on $\Delta_1$ and $\Delta_{2(day)}$.

Suspicious screening process (Fig.2: step (3)) allows us to choose the most relevant subset of whole datasets and moreover we only need to perform the clustering algorithm one time to determine the suspicious group. Briefly, let $V$ be the set of transactions datasets aggregated by $\Delta_1$ and $\Delta_2$:

$$V = \bigcup_{i=1}^{n} v_i$$

and this process tries to find a relevant subset $V' \subset V$, $V'$ is formally defined as:

$V' = \{v_i (\Delta_{1i}, \Delta_{2i}) \in V / s \leq \Delta_{1i} \leq 1 \wedge S \leq \Delta_{2i} \leq 1, s, S \in R \wedge s \in [0..1] \wedge S \in [0..1]\}$

More details of this process can be found in [16]. Fig. 4, for example shows the clustering results of $V'$ from the datasets $V$ of the fund SK where $s = S = 0.4$.

These outputs will be then fed into a neural network (back propagation based) for training on suspicious and non-suspicious cases. We build also a decision tree on training results conditional on time (Fig.2: step (5)). Knowledge are then stored in a knowledge repository that assists the AML experts making a decision (Fig.2: step (6)). The rest of the transaction monitoring process can be resumed as following: in order to investigate one case, its transactions are firstly placed in a suitable period level (day, week, month, etc., Fig.2: step (7)). At the time $t$ on the day $d$, an AML expert investigates the customer X's transactions then its $\Delta_1$, $\Delta_2$ of this day $d$ and its relevant $\Delta_1$, $\Delta_2$ in this period are calculated. For example, if the choosing period is "month" then $(\Delta_1, \Delta_2)_{day}$ of the day $d$ and $(\Delta_1, \Delta_2)_{week}$, $(\Delta_1, \Delta_2)_{month}$ for the week and the month respectively that contain $d$ are also calculated.

Next, these parameters will be fed in a trained neural network (Fig.2: step (8)) to determine the suspicious degree of each level i.e. at the day level, the week level and the month level as in the previous example. AML expert uses these suspicious degrees to determine suspicious cases and carries out further investigation (e.g. redemption type, nationality, country of residence, etc.) on these cases (Fig.2: (10)).

## V. EVALUATION AND ANALYSIS

We evaluate our approach using transactions from funds administered by **BEP bank**[2] with two millions transaction records. The testing environment is Windows XP with 2Gb RAM, 3.4GhZ Intel Dual Core and .NET platform. Transaction datasets are divided into two groups according to two kinds of investors: individual and corporate because they are relatively different in their investment behaviour. Another filtering on our datasets is the mapping error cleaning as we mentioned in Section III.A. Besides, the parameters set up for training neural networks are: training cycles = 5000, learning rate = 0.25, number of layers = 3.

Fig.4, Fig.5 and Fig.6 show the clustering results of the fund SK based on $\Delta_1$ and $\Delta_2$ by day, by week and by month respectively and clustering times is variety from 0.1s to 0.3s. By observing these results, we recognise that elements in Cluster4 (Fig.4), Cluster2 (Fig.5) and Cluster4 (Fig.6) are obtained high value in $\Delta_1$ and $\Delta_2$ and their values will be used as suspicious cases.

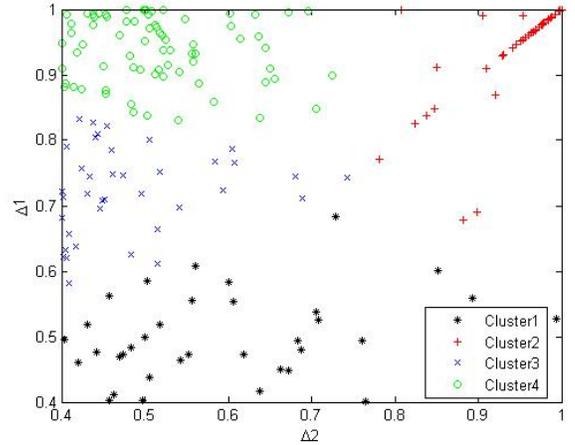

Figure 5. Clustering of Fund SK datasets based on $\Delta_1$ and $\Delta_{2(week)}$.

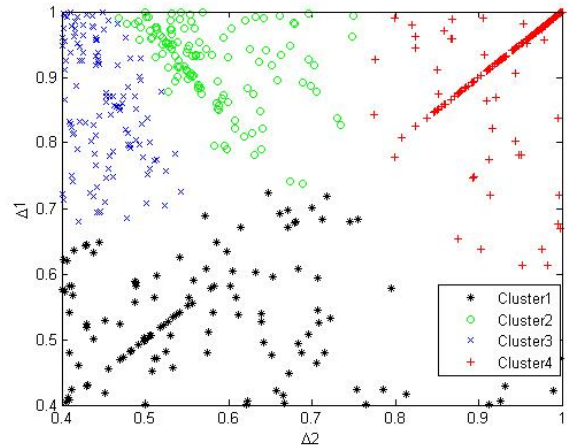

Figure 6. Clustering of Fund SK datasets based on $\Delta_1$ and $\Delta_{2(month)}$.

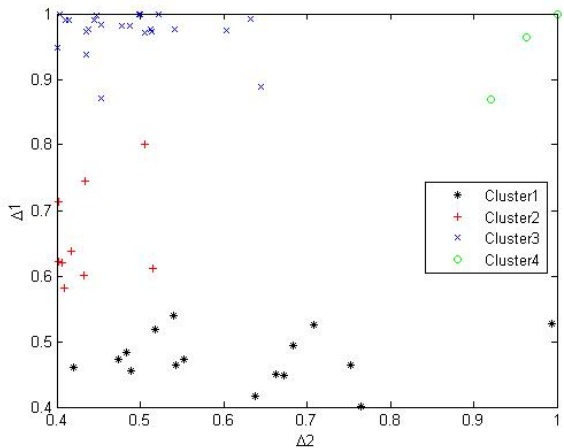

Figure 4. Clustering of Fund SK datasets based on $\Delta_1$ and $\Delta_{2(day)}$ where $s = S = 0.4$.

---

[2] Real name of the bank can not be disclosed because of confidential agreement of the project

Experience of training for the fund SK for corporate customers is shown in the Table II.

TABLE II. EXPERIENCE OF TRAINING SUSPICIOUS CASES OF THE FUND SK FOR CORPORATE CUSTOMERS

| Period | Number of suspicious records | Number of unsuspicious records | Training time in total |
|---|---|---|---|
| Day | 7 | 2000 | 28s 482ms |
| Week | 47 | 2000 | 32s 151ms |
| Month | 179 | 2000 | 32s 918ms |

In this case, we take 100% of suspicious cases (obtained high value in ($\Delta_1$, $\Delta_2$) and returned by the clustering process) for training. In the unsuspicious case, we take from 5% to 10% of population for training. Table III shows some examples of suspicious degree detected by our neural networks. Total running time for this process is ~ 0.51 second for ~45000 records. Besides, the third case in the Table III can be explained as on 31/05/00 the customer 3310 redeems an amount of 97% of his subscription in a the last few days and this amount is moreover about 80% of his total values of shares. This case is clearly a suspicious case. These suspicious cases are then investigated further and most of them have exchange transactions i.e. one can redeem his/her entire share from one sub-fund and invest into another sub fund. Both two sub-funds are in the same investment fund. After the refinement process, the real suspicious cases were approximate 5. This is consistent with reports from **BEP**'s bank by using a manual approach that takes more than a week to detect.

TABLE III. EXPERIENCE OF INVESTIGATING SUSPICIOUS CASES OF THE FUND SK FOR CORPORATE CUSTOMERS

| CID | Date | $\Delta_1$; $\Delta_2$(day) | Suspicious degree |
|---|---|---|---|
| 515 | 21/09/99 | 0.98;0.83 | 0.99 |
| 1074 | 22/01/01 | 0.0019;0.01 | 0.00072 |
| 3310 | 31/05/00 | 0.9732;0.79 | 0.99 |

## VI. CONCLUSIONS AND FUTURE WORKS

In this paper, we have presented a DM solution for analysing transactions in an investment bank to detect ML. In our approach, we determined first of all, the important factors for investigating ML in the investment activities. Next, we proposed an investigating process based on clustering and neural network to detect suspicious cases in the context of ML. We also applied heuristics such as suspicious screening to improve the running time. From our experimental results obtained on the greatest fund of **BEP**'s transaction datasets, we can conclude that our approach is promising and it satisfies the needs of the AML unit. It can also improve significantly the performance from our previous solution in terms of running time.

Experimental results on real-platforms of **BEP** for all transaction datasets are also being produced and these will allow us to test and evaluate the robustness of our approach. We are currently working on improving the learning process to tackle the problem of very large datasets.